\documentclass[prl,twocolumn,showpacs]{revtex4}

\usepackage{graphicx}
\usepackage{amsmath}
\usepackage{amssymb}

\begin{document}
\def\a{\alpha}
\def\b{\beta}
\def\e{\varepsilon}
\def\d{\delta}
\def\l{\lambda}
\def\m{\mu}
\def\t{\tau}
\def\n{\nu}
\def\o{\omega}
\def\r{\rho}
\def\S{\Sigma}
\def\G{\Gamma}
\def\D{\Delta}
\def\O{\Omega}

\def\ra{\rightarrow}
\def\ua{\uparrow}
\def\da{\downarrow}
\def\pd{\partial}
\def\bk{{\bf k}}
\def\bp{{\bf p}}
\def\bn{{\bf n}}

\def\be{\begin{equation}}\def\ee{\end{equation}}
\def\bea{\begin{eqnarray}}\def\eea{\end{eqnarray}}
\def\nn{\nonumber}
\def\lb{\label}
\def\pref#1{(\ref{#1})}


\title{Nodal quasiparticles in doped d-wave superconductors: self-consistent
T-matrix approach}

\author{V.M.~Loktev$^{1}$}
\author{Yu.G.~Pogorelov$^{2}$}
\affiliation{$^1$Bogolyubov Institute for Theoretical Physics, Kiev-143, Ukraine\\
        $^2$CFP/Departamento de F\'{i}sica, Universidade do Porto, 4169-007 Porto,
Portugal}

\date{\today }

\begin{abstract}
A comparative analysis has been done of the formerly established two
self-consistent solutions for the density of quasiparticle states in doped
d-wave superconductors, having strikingly different and disputed behavior
near the Fermi energy. One of them (1) remains finite in this limit, while
the other (2) tends to zero. To resolve this discrepancy, the known
Ioffe-Regel criterion for band states, widely used for doped semiconductors,
was applied to these solutions. It is shown that both them are valid in limited
and different energy regions, where the corresponding quasiparticles are
weakly damped. In particular, density of states of nodal quasiparticles near
the Fermi level is provided by the (2) solution, while the (1) only applies
far enough from this level.

\end{abstract}




\maketitle

The self-consistent T-matrix approximation (SCTMA \cite{bay}, or FLEX method 
\cite{bick}) is extensively used for description of quasiparticles and their
density of states (DOS) in disordered crystals, in particular, in the doped
high-Tc superconducting cuprates. Its advantage consists in relative
simplicity and transparency for numeric calculations. However, this
simplicity can be sometimes mischievous if the solutions are used without
limitations on their validity. In fact, a number of controversies emerge
when comparing some SCTMA results \cite{zieg} with those of other approaches 
\cite{ners},\cite{atk}, which suggest the need for a more detailed
substantiation of the method. Perhaps the central point in this discussion
is now the question about the quasiparticle DOS $\r (\e)$ close to the center 
of superconducting gap, that is at $\e \ra 0$, where the DOS of clean d-wave 
superconductor vanishes as $\r_{d}(\e )\propto \mathrm{const}\cdot|\e|$. The
theoretical predictions for its behavior in presence of impurity scattering
include: \textit{i}) various kinds of tendency to zero \cite{ners}, \cite{atk}, 
\cite{sent}, \cite{lp}, \textit{ii}) tendency to a constant value \cite{gor}, 
\cite{lee}, and even \textit{iii}) to infinity \cite{pep}. Such extreme
ambiguity motivates us to reconsider this situation in a more general context 
of the theory of elementary excitations in disordered systems \cite{lgp}.

\begin{figure}
\centering{
\includegraphics[width=6.cm, angle=0]{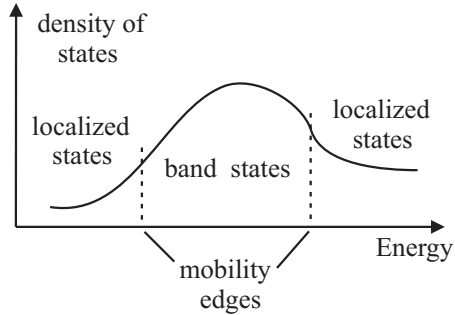}}
\caption{General Mott picture of spectrum of elementary excitations in a 
disordered system.}
\lb{fig1}
\end{figure}

There is a consensus on that the single-particle excitations in disordered
sytems can be either of band (extended) type or localized type \cite{mott},
both types forming certain continuous regions of spectrum separated by the
so-called mobility edges $\e _{c}$ (Fig. \ref{fig1}). The extended states can 
be approximately described by the wavevector $\bk$, through a dispersion law 
$E_{\bk}$ and a broadening $\G_{\bk}$, as far as the Ioffe-Regel criterion (IRC) 
\cite{ir} for the mean free path $\ell $ and the wavelength $\l$ holds: 
\be
\ell \gg \l ,\ \ \ \ \ \  \mathrm{or}\ \ \ \ \ \ k \left| \pd
E_{\bk}/\pd \bk \right| \gg \G_{\bk}.
\lb {eq1}
\ee
Then the real and imagnary parts of self-energy $\S_{\bk}$ in 
the disorder averaged Green function (GF) $G_{\bk}=(\e - \e_{\bk}-\S_{\bk})^{-1}$ 
permit to define: $E_{\bk}=\e_{\bk}-\mathrm{Re}\S_{\bk}\left( E_{\bk}\right)$, 
$\G_{\bk}=\mathrm{Im}\S_{\bk}\left( E_{\bk}\right)$ \cite{bb}, and the 
modified DOS $\r(\e)=\pi^{-1}\mathrm{Im}G \approx \r_{0}(\e)/
\left(\pd E_{\bk}/\pd\e_{\bk} \right)_{E_{\bk}=\e}$ ($\e_{\bk}$ and $\r_{0}(\e)$ 
being respectively the dispersion law and DOS in pure crystal). The condition $\ell 
\sim \l$ is reached when $E_{\bk}$ approaches the mobility edges, then the very notion 
of self-energy correction to the initial band spectrum $\e_{\bk}$ ceases to make sense 
and the averaged properties of localized states are only described by their DOS $\r(\e)$.

In the case of disorder due to fixed impurity perturbation $V_{\mathrm{L}}$ at random
sites $\bp$ (Lifshitz model) for the normal metal quasiparticles, the
modification of band spectrum can involve new specific features like local
or resonance levels \cite{lif}, and there are various ways to expand the
self-energy in groups of interacting impurity centers \cite{ilp}, analogous
to the classical Ursell-Mayer group expansions (GE) for statistical sum \cite
{hill}. For instance, the so-called fully renormalized GE \cite{iv} reads: 
\bea
\S_{\bk}&=&\sum_{\bp}\frac{V_{\mathrm{L}}}{1-GV_{\mathrm{L}}}\left[1+ \right. \notag \\
&&\left. + \sum_{\bp^\prime \neq \bp} \frac{A_{\bp,\bp^\prime}\mathrm{e}^{i\bk\left(\bp^\prime - \bp\right)} 
- A_{\bp,\bp^\prime} A_{\bp^\prime,\bp}}
{1-A_{\bp,\bp^\prime}A_{\bp^\prime,\bp}}+\ldots \right],  \lb{eq2}
\eea
where $A_{\bp,\bp^\prime}=V_{\mathrm{L}}G_{\bp,\bp^\prime}/(1-GV_{\mathrm{L}})$ includes $G_{\bp,\bp^\prime}=
N^{-1}\sum_{\bk^\prime \neq \bk} \mathrm{e}^{i\bk^\prime\left( \bp-\bp^\prime \right)} G_{\bk}$ and $G=G_{\bp,\bp}$. 
Other types of GE can differ from Eq.\ref{eq2} either in the structure of next to unity terms and in the
degree of renormalization of $G$ and $A$ functions present in them. The relevant 
expansion parameter is not simply the impurity concentration $c=\sum_{\bp}N^{-1}$ 
(supposedly small, $c\ll 1$) but the ''gas parameter'' $c\sum_{\mathbf{n}\neq 0} 
A_{0,\bn}^{2}$ for the ''non-ideal gas'' of impurities with effective interaction described 
by the (energy dependent) functions $A_{\bp,\bp^\prime}$. Hence the convergence of
the series (\ref{eq2}) turns also energy dependent, reflecting the division between the 
above referred types of states. It can be shown that this convergence is equivalent to 
validity of the IRC \cite{is}. Within the energy domain of convergence, the self-energy 
can be approximated by Eq.\ref {eq2} with only unity term retained in the brackets: 
$\S_{\bk}\approx \S =cV_{\mathrm{L}}/(1-GV_{\mathrm{L}})$, the momentum-independent
SCTMA form. But beyond this domain, the SCTMA does not make sense and a better description 
of DOS is obtained with GE's, different from Eq.\ref{eq2}. Below we check the fulfillment 
of IRC for nodal quasiparticle states in a d-wave superconductor with dopants and conclude 
on the validity of the SCTMA solutions.

Let us start from the most common model Hamiltonian for this problem:

\bea
H&=&\sum_{\bk}\psi_{\bk}^{\dagger}\left(\xi_{\bk}\widehat{\t}_{3}+\D_{\bk}\widehat{\t}_{1}\right) \psi_{\bk}  \nn \\
&&- \frac{V_{\mathrm{L}}}{N}\sum_{\bk,\bk_{\prime},\bp}\mathrm{e}^{i(\bk-\bk^\prime)\bp}
\psi _{\bk}^{\dagger}\widehat{\t}_{3}\psi_{\bk^\prime},  \lb{eq3}
\eea
Here the Nambu spinors $\psi _{\bk}^{\dagger }=\left( c_{\bk,\ua}^{\dagger},
c_{-\bk,\da}\right) $ include Fermi operators of normal quasiparticles with the simplest 
2D dispersion law $\xi_{\bk}=2t\left( 2-\cos ak_{x}-\cos ak_{y}\right) -\m$ in square
lattice, approximately constant DOS $\r(\e)\approx \r_{0}=4/\left( \pi W\right) $ where 
$W=8t$ is the bandwidth, and chemical potential $\m$; $\widehat{\t}_{j}$ are the Pauli 
matrices. The d-wave gap function is $\D_{\bk}=\D \theta \left( \e_{D}^{2}-\xi _{\bk}^{2}\right) 
\cos 2\varphi _{\bk}$, where the ''Debye'' energy $\e_{D} \gg \D$, and $\varphi _{\bk}=\arctan k_{y}/k_{x}$ 
defines the nodal lines $k_{x}=\pm k_{y}$. The Lifshitz perturbation term in Eq.\ref{eq3} produces scattering 
of quasiparticles, modelling the impurity effect of dopants.

The relevant GF is a Nambu matrix $\widehat{G}_{\bk}=\left\langle \left\langle \psi _{\bk}|
\psi _{\bk}^{\dagger }\right\rangle \right\rangle $ with matrix elements being Fourier transformed two-time GF's: 
\be
\left\langle \left\langle a|b\right\rangle \right\rangle _{\varepsilon
}=i\int_{0}^{\infty }\mathrm{e}^{i(\mathbf{\varepsilon +}i0)t}\left\langle
\left\{ a\left( t\right) ,b\left( 0\right) \right\} \right\rangle dt \nn
\ee
where $\left\langle \ldots \right\rangle $ is the quantum statistical average
with the Hamiltonian \ref{eq3} and $\left\{ . , . \right\} $ is the
anticommutator of Heisenberg operators. In analogy with the above scalar GF 
$G_{\bk}$ for normal quasiparticles, the general solution for this matrix is 
\be
\widehat{G}_{\bk}=\left( \e -\m -\xi_{\bk}\widehat{\t}_{3} - \D_{\bk}\widehat{\t}_{1} - 
\widehat{\S}_{\bk}\right)^{-1}.  \lb{eq4}
\ee
All the impurity effects are now accounted for by a GE for the self-energy
matrix $\widehat{\S}_{\bk}$ \cite{po1} (cf. Eq.\ref{eq2}): 
\bea
\widehat{\S}_{\bk} &=&\frac{1}{N}\sum_{\bp} \widehat{V}\left( 1 - \widehat{G}\widehat{V} \right)^{-1}
\left\{ 1 + \right. \notag  \\
&& + \sum_{\bp^\prime \neq \bp} \left[ \widehat{A}_{\bp,\bp^\prime} \mathrm{e}^{i \bk \left( \bp^\prime - \bp \right)} 
- \widehat{A}_{\bp,\bp^\prime} \widehat{A}_{\bp^\prime,\bp} \right]\times \notag \\ 
&& \left. \times \left[ 1 - \widehat{A}_{\bp,\bp^\prime} \widehat{A}_{\bp^\prime,\bp} \right]^{-1} + \ldots \right\} .  
\lb{eq5}
\eea
Here the matrices: $\widehat{V}=V_{\mathrm{L}}\widehat{\t}_{3}$, $\widehat{G}=N^{-1}\sum_{\bk}\widehat{G}_{\bk}$, and 
\be
\widehat{A}_{\bp,\bp^\prime}=N^{-1}\sum_{\bk^\prime \neq \bk} \mathrm{e}^{i\bk^\prime 
\left( \bp - \bp^\prime \right)} \widehat{G}_{\bk^\prime} \left( 1- \widehat{G} \widehat{V}\right)^{-1}, \nn
\ee
and some additional restrictions are imposed on summation in momenta in the
products like $\widehat{A}_{\bp,\bp^\prime} \widehat{A}_{\bp^\prime,\bp}$, resulting 
from a specific procedure of consecutive elimination of GF's in the infinite chain of
coupled Dyson equations. There are possible different such procedures and,
respectively, different types of GE \cite{ilp}. Generally, GE's are only
asymptotically convergent and the best choice between them is determined by
their convergence range with respect to energy $\e $.

The conditions for convergence of different GE's were studied in detail for
a number of types of elementary excitations in crystals with impurities \cite
{ip},\cite{lp}, and this permitted to establish certain general criteria for
the corresponding characteristic regions of spectrum. In particular, the
region of band states is best described by the so-called fully renormalized
GE which ceases to converge at approaching the mobility edges where IRC, Eq.\ref{eq1}, 
fails. For the GE, this is expressed by the tendency of all its
terms, next to the unity in curled brackets in Eq.\ref{eq5}, to become $\sim$1.

Alike the above mentioned scalar case, the SCTMA just corresponds to the fully
renormalized GE, restricted to only its first term. Hence it is only justified when 
IRC holds. Bearing this in mind, let us analyze the SCTMA solutions in the vicinity 
of the Fermi energy for the system, described by the Hamiltonian \ref{eq3}.

Then, using the $\bk$-independent SCTMA self-energy $\widehat{\S}=\widehat{V}
\left( 1-\widehat{G}\widehat{V}\right) ^{-1}$ and following the procedure of Refs. 
\cite{lp2},\cite{lp}, one can arrive at the explicit average local GF: 
\be
G=\frac{1}{2}\mathrm{Tr}\widehat{G}=\widetilde{\e }\r_{0}\left[ 
\frac{1}{\widetilde{\m}}+ \frac{2}{\sqrt{\widetilde{\e}^{2}-\D^{2}}}\mathrm{K}
\left( \frac{\D^{2}}{\D^{2}-\widetilde{\e}^{2}}\right) \right] .  \lb{eq6}
\ee
In Eq.\ref{eq6}, the renormalized energy $\widetilde{\e}=\e -\S \left( \widetilde{\e}\right)$ 
includes the scalar value $\S =\mathrm{Tr}\widehat{\S}/2$, the parameter $\widetilde{\m}=\m
(1-\m \r_{0}/2)$, and $\mathrm{K}$ is the 1st kind full elliptic integral. Having the 
explicit relation \cite{lp} $\Sigma =c\widetilde{V}^{2}G/(1-\widetilde{V}^{2}G^{2})$ between 
the self-energy and GF, where the renormalized perturbation parameter $\widetilde{V} = V_{\mathrm{L}} 
/ [1+V_{\mathrm{L}}\r_{0}\ln (1-\widetilde{\m}/\m)]$, one obtains the self-consistent equation for 
$G$ which can be solved in principle numerically. Since the analytic structure of Eq.\ref{eq6} 
involves singular points in the complex $G$ plane (including essential singularity of the 
$\mathrm{K}$-function), it possesses multiple solutions. The physical solutions among them should be 
then selected by IRC, as a necessary condition for SCTMA validity.

\begin{figure}
\centering{
\includegraphics[width=9.cm, angle=0]{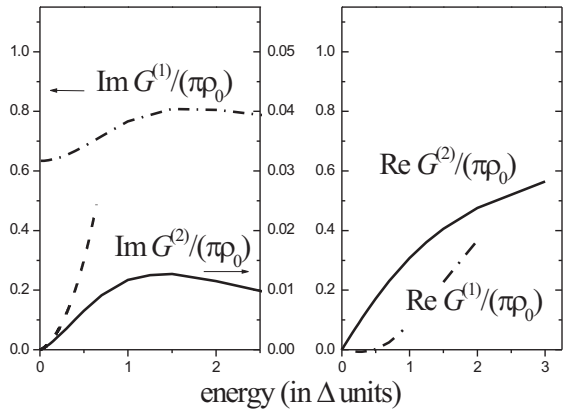}}
\caption{Real and imaginary parts of self-consistent Green functions satisfying 
Eq.\ref{eq6}. Solid lines are for $G^{(2)}$ solution with the asymptotics, Eq.\ref{eq7} (dashed line), 
and dash-dotted lines for $G^{(1)}$ solution.}
\label{fig2}
\end{figure}

The analysis turns most transparent in the important limit $\e \ra 0$ (related to the 
Fermi energy). There are two characteristic solutions \cite{foot} of Eq.\ref{eq6} 
in this limit. One of them, $G\left( \varepsilon \right) = G^{\left( 1\right) }\left(\e 
\right) $, tending to a constant imaginary value, $G^{\left(1\right)}\left( \e \ra 0\right) 
\ra i\cdot \mathrm{const}$, was first obtained by Gor'kov and Kalugin in the Born scattering 
limit \cite{gor} and then by P.A. Lee in the unitary scattering limit \cite{lee}. Later on, 
it was repeatedly reproduced by various numerical techniques \cite{atk} and hence believed to 
be the unique SCTMA solution. Neverthless, it was shown recently by the authors \cite{lp} that 
another solution exists, $G\left( \e \right) = G^{\left( 2\right) }\left(\e \right) $, with low 
energy asymptotics: 
\be
G^{\left( 2\right) }\left( \e \right) \approx \frac{\e }{c\widetilde{V}^{2}}\left[ 1-
\frac{\D}{ \pi c\widetilde{V}^{2}\r_{0}\ln \left( 2 i \pi c\widetilde{V}^{2}\rho _{0}/\e \right)} \right] ,
\lb{eq7}
\ee
which tends to zero with $\e $. The behavior of real and imaginary parts of the two solutions 
in function of energy for a particular choice of parameters: $c=10\%$, $V_{\mathrm{L}}=0.3$ eV, $W=2$ eV, and $\D =30$ meV, 
is shown in Fig.\ref{fig2}. Notice that at low energies, $\e \ll \D $, the solution $G^{\left( 1\right) }$ is 
dominated by the above mentioned imaginary constant, presented as $i\pi c\r_{0}g_{0}$ where $g_{0}\lesssim 1$ is a root of 
a certain transcendental equation \cite{lp}. In contrary, the tendency of $G^{\left( 2\right) }$ to zero is characterized 
by the progressive domination of its real part. The renormalized dispersion law $\widetilde{E}_{\bk}$ (as far as the 
condition \ref{eq1} holds) is given by the common equation \cite{bb} 
\be
\widetilde{E}_{\bk}-\mathrm{Re}\S \left( \widetilde{E}_{\bk}\right) =E_{\bk},  
\lb{eq8}
\ee
where $E_{\bk}=\sqrt{\xi _{\bk}^{2}+\D_{\bk}^{2}}$ is the non-perturbed superconducting dispersion law and $\mathrm{\S }$ is 
specified for particular $G^{\left( j\right) }$, $j=1,2$. Then the IRC is written down as: 
\be
\left( \bk-\bk_{0} \right) \cdot \mathbf{\nabla }_{\mathbf{k}}\widetilde{E}_{\bk} 
\gg \G _{\bk}, \nn
\ee
near a nodal point $\bk_{0}$ where a nodal line crosses the Fermi surface.

The low energy asymptotics of Eq.\ref{eq8}, corresponding to the $G^{\left(2\right) }$ solution, 
Eq.\ref{eq7}, is: $\widetilde{E}_{\bk}^{\left(2\right) } \approx (\pi c\widetilde{V}^{2}\r_{0}/\D)
E_{\bk}\ln (2\D/E_{\bk})$, and with the related damping $\G_{\bk}^{\left( 2\right) }=\mathrm{Im}\S^{\left( 2\right) }
\left(\widetilde{E}_{\bk}^{\left( 2\right) }\right) \approx E_{\bk}/\ln (2\D/E_{\bk})$, we 
arrive at the condition: 
\be
E_{\bk} \ll \D \exp \left( -\sqrt{\frac{\D }{\pi c\widetilde{V}^{2}\r_{0}}}\right) ,  
\lb{eq9}
\ee
which defines a narrow enough vicinity of the Fermi energy where this
solution makes sense.

Applying the same treatment to the $G^{\left( 1\right) }$ solution, which
formally defines the low energy dispersion law $\widetilde{E}_{\bk}^{\left( 1\right) } 
\approx E_{\bk}$ and the damping $\G_{\bk}^{\left( 1\right) }=\mathrm{Im}
\S ^{\left( 1\right)}\left( \widetilde{E}_{\bk}^{\left( 1\right) }\right) \approx 
\pi c \widetilde{V}^{2}\r_{0}g_{0}$, we obtain the condition 
\be
E_{\bk} \gg \pi c\widetilde{V}^{2}\r_{0}g_{0},  
\lb{eq10}
\ee
so that this solution is valid far enough from the nodal points, where it
provides also a correct limit of pure d-wave DOS. However, this solution is
clearly eliminated near the nodal point. Thus we come to the conclusion that
the only SCTMA solution, valid in the close vicinity of the Fermi energy, is
that given by Eq.\ref{eq7}. A physical consequence of vanishing DOS at $\e \to 
0$ for this solution is that the much disputed conjecture of universal electric 
and thermal conductivity \cite{lee}, \cite{durst} turns impossible. Nevertheless, 
if the validity range for the $G^{\left(2\right)}$ solution, Eq.\ref{eq9}, is very 
narrow, these conductivities, as far as being defined by the $G^{\left(1\right)}$ 
solution, can display an apparent tendency to those universal values.

Notably, the two estimates, Eqs.\ref{eq9},\ref{eq10}, do not necessarily
assure the overlap between the two validity regions, so that for $\pi c 
\widetilde{V}^{2}\r_{0} \gg \D $ there can exist some intermediate
energy range where neither of SCTMA solutions applies. This range roughly
corresponds to the broad linewidth of the known impurity resonance $\e_{res}$ 
\cite{lp} where DOS cannot be rigorously obtained even with use of the next 
terms from GE, Eq.\ref{eq6}, though some plausible interpolation is possible between 
the two SCTMA asymptotics.

Finally it is worthwhile to notice that other known non-perturbative solutions for d-wave
disordered systems with DOS vanishing at $\e \ra 0$ as a certain power law: $\r 
\left( \e \right) \sim \e^{\a}$ \cite{ners},\cite{sent}, also have to satisfy IRC since 
they use field theoretic approach, only compatible with band-like states. But it can
be easily shown that this criterion can be only fulfilled for such DOS if the power 
is $\a >1$, while the reported values are $\a =1/7$ \cite{ners} and $\a =1$ \cite{sent}.

In fact, let the renormalized radial dispersion law (in the low energy limit) behave as 
$\widetilde{\xi }_{k}\sim (k-k_{\mathrm{F}})^{\n } \propto \xi ^{\n}$ with certain $\n >0$, 
then the simplest estimate for d-wave DOS is 
\bea
\r \left( \e \right) &\propto &\e \int_{0}^{\D} d\eta \int d\xi \d \left( \e^{2} - 
\widetilde{\xi }^{2} - \eta^{2}\right) \nn \\
&\propto &\e \int_{0}^{\e }\frac{d\eta }{\left(
\e^{2}-\eta ^{2}\right) ^{\n -1/2}}=\frac{\sqrt{\pi }\G \left( 3/2-\n \right)}
{4\n \Gamma \left( 2-\n \right) }\e^{3-2\n}, \nn
\eea
that is $\a =3-2\n$. In the considered field models, DOS defines the
quasiparticle broadening $\G_{\bk}=u^{2}\r \left( \widetilde{\xi }_{k}\right) $, with a 
disorder parameter $u$. Then the criterion, Eq.\ref{eq1}, is reformulated as 
\be
\xi \frac{d\widetilde{\xi }}{d\xi }\gg u^{2}\r \left( \widetilde{\xi}_{k}\right) , \nn
\ee
leading to the condition $\xi ^{\n }\gg \mathrm{const\cdot }\xi ^{\n \left( 3-2\n \right)}$, 
and in the limit $\xi \ra 0$ this is only possible if $3-2\n >1$, that is $\a >1$.

So, the above considerations essentially restrict possible candidate
solutions for quasiparticle spectrum in the disordered d-wave superconductor
and in fact suggest Eq.\ref{eq7} as the only known consistent low energy
solution for the problem.

We thank Hans Beck, Valery Gusynin, Vladimir Miransky, and Sergei Sharapov
for valuable and stimulating discussions. We also acknowledge the Swiss
Science Foundation for the partial support of this research (the SCOPES
grant 7UKPJ062150.00/1) and Neuchatel University for kind hospitality.


\begin{thebibliography}{99}
\bibitem{foot} It can be shown that, besides these two, formally there are also 
other solutions, but all they fail to satisfy IRC.
\end{thebibliography}
\end{document}